\documentclass[11pt,a4paper,onecolumn]{revtex4}    
\usepackage{graphicx}
\begin{document}           
\preprint{0th version}

\title{Compensation of eddy--current--induced magnetic field transients in a MOT}

\author{C. L. Garrido Alzar}
\email{leonardo@nbi.dk}
\affiliation{QUANTOP, Danish National Research Foundation Centre of
  Quantum Optics, Niels Bohr Institute, Blegdamsvej 17, DK-2100 Copenhagen \O, Denmark.}
\author{P. G. Petrov}
\affiliation{QUANTOP, Danish National Research Foundation Centre of
  Quantum Optics, Niels Bohr Institute, Blegdamsvej 17, DK-2100 Copenhagen \O, Denmark.}
\author{D. Oblak}
\affiliation{QUANTOP, Danish National Research Foundation Centre of
  Quantum Optics, Niels Bohr Institute, Blegdamsvej 17, DK-2100 Copenhagen \O, Denmark.}
\author{J. H. M\"uller}
\affiliation{QUANTOP, Danish National Research Foundation Centre of
  Quantum Optics, Niels Bohr Institute, Blegdamsvej 17, DK-2100 Copenhagen \O, Denmark.}
\author{E. S. Polzik}
\affiliation{QUANTOP, Danish National Research Foundation Centre of
  Quantum Optics, Niels Bohr Institute, Blegdamsvej 17, DK-2100 Copenhagen \O, Denmark.}

\date{\today}

\begin{abstract}
The design and implementation of a current driver for the quadrupole coil of a magneto-optical 
trap are presented. The control and the power stages of the driver,
both based on a push-pull configuration with high-speed transistors, are
separated, allowing for a protection of the control
electronics. Moreover, the coil current can be set by an
analog voltage, a feature that makes possible both fast and adiabatic
switching of the quadrupole magnetic field. In order to compensate magnetic field transients
induced by eddy currents, the driver allows a quick reversal of the quadrupole
coil current providing in such a way the required magnetic flux
compensation. From a fit to the measured magnetic field
transients, we have extracted the decay constants which confirm the
efficiency of the compensation. Furthermore, we measured the influence
of eddy currents on a sample of cold caesium atoms, testing at the
same time the ultimate performance of the driver for our applications. The
results show that the implemented electronic circuit is able to reduce
substantially the transient effects with switching times of the magnetic field below 100 $\mu$s.
\end{abstract}

\maketitle

\section{Introduction}
Magnetic field gradients play a central role in the preparation of cold atomic 
samples in magneto-optical traps (MOT). However, for many experiments involving 
cold atoms the presence of magnetic field gradients is unwanted. This means that, 
following the preparation of a cold atomic sample in a MOT, the time interval that 
we have at our disposal to perform an experiment is, to some extent, determined 
by the time required to turn off completely the current flowing through the coils that 
produce the quadrupole magnetic field of the MOT. More precisely, is the time needed to 
bring the magnetic field gradient to zero that sets the ultimate limit.

The current flowing in the coils can be brought to zero in several microseconds using 
the common technique that employs a suppression diode for switching inductive 
loads~\cite{theart}. To bring the magnetic field to zero is more difficult since the 
fast change of the magnetic flux, produced by the coil current, induces eddy currents in the metallic parts 
adjacent to the coils. These currents will produce a time and spatially varying magnetic 
field that dies out typically in several milliseconds, depending on
the experimental setup.

For a MOT, the negative effects due to magnetic field transients (and
their associated gradients) generated by eddy currents 
are seen immediately. This parasitic transient washes out the sample of cold atoms disturbing, 
for instance, experiments like loading of optical dipole traps. In
particular, in our experimental setup the metallic environment 
around the coils producing the magnetic quadrupole field is different or asymmetric for the 
two coils and consequently, the trap center moves during the switching transient accelerating the atoms, an 
effect which is deleterious for transferring the atoms into the optical trap. Yet in another 
field of research as magnetic resonance imaging (MRI)~\cite{MRI1st}, the eddy--current--induced 
gradients reduce the signal to noise ratio degrading the quality of the obtained images.

It is possible to find different proposals for the compensation of
eddy currents. A method 
based on gradient with pre-emphasis correction~\cite{correc} is employed in, for example, MRI~\cite{MRI} and 
particle storage rings~\cite{rings}, but it requires the monitoring of the magnetic field in 
order to implement a closed-loop compensation. Also for MRI and
nuclear magnetic resonance (NMR), radiofrequency shielding of the gradient coils has been
developed~\cite{shield}. Another more interesting, for our purposes, 
compensation scheme has been recently implemented by Dedman \textit{et al.} for fast switching 
of magnetic fields in a MOT~\cite{dedman}. In their proposal two power MOSFET are used to 
switch off or reverse the polarity of the current flowing through the coils, which allows switching times for the magnetic field of around 350 $\mu$s. However, it is our belief that 
the performance of the circuit can be improved in two main directions. In terms of protection, 
it is important to avoid the use of high voltage power supplies that can seriously damage the 
control voltage source (a PC card, for example) in case of failure of the transistors. And, on 
the other hand, to allow an analog control of the amount of current fed to the coils, a 
relevant feature for applications like for example the preparation of cold samples in the Cs 
atomic fountain clock~\cite{adia}, where it is necessary to combine fast as well as adiabatic 
switching of the quadrupole magnetic field.

Here we propose a current driver that combines high-speed switching transistors on the 
control stage of the circuit and power switching bipolar transistors to handle the 
coils' current. Such a design allows the use of stable low voltage (max. 10 Volts) low 
current (max. 10 Amps) standard laboratory power supplies, fast and adiabatic switching of 
magnetic fields, and is as practical to implement as the MOSFET solution. We 
also provide the design considerations to help the interested reader in choosing the proper 
component specifications for her/his particular application.

\section{Fast switching current driver design}
Usually, the magnetic field $\vec{B}(t)$ transient properties are measured with a
pick-up coil which generates a voltage
\begin{equation}
 V_p(t)=-N\frac{d\Phi(t)}{dt} \ ,
\label{picvolt}
\end{equation}
being $N$ and $\Phi(t)=\int \vec{B}(t)\vec{dS}$ the number of turns of the coil and the
magnetic flux sensed by it, respectively. Therefore, choosing the coil
area $S_c$ such that the field across it is homogeneous, $V_p(t)$ will
provide a direct measurement of $\frac{dB(t)}{dt}$. Qualitatively, the
ideal (desired) magnetic field switching waveform can be represented
by a square pulse and, in this case, the voltage measured by the
pick-up coil can be mathematically described by the expression 
\mbox{$V_p(t)=-N S_c B_{eff}(\delta(t_1)-\delta(t_2))$}, where
$B_{eff}$, $\delta(t)$, $t_1$, and $t_2$ are the effective
magnetic field amplitude, the Dirac delta function and
the time instants at which the transitions take place, respectively. Nevertheless, the actual magnetic field
transient waveform shows exponential transitions characterized by a
time constant determined mainly by two effects: the
Lenz effect and the existence of eddy currents induced in the metallic
structures close to the coils producing the quadrupole magnetic field.

\begin{figure}[ht]
\begin{center}
\includegraphics*[width=0.35\textwidth]{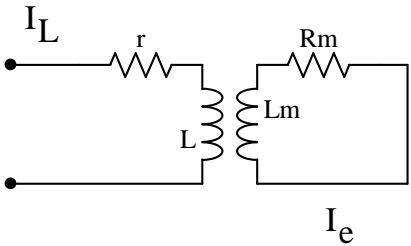}%
\end{center}
\caption{Modelling of induced eddy currents.}
\label{fig:model}
\end{figure}

Just as a reminder, the influence of eddy currents can be studied with
the simple electric circuit model presented in Fig.~\ref{fig:model}. The elements $L$ and $r$ represent the
inductance and the wire resistance of the MOT coils, respectively. The metallic environment of the coils can be described
by the inductance $L_m$ and the resistance $R_m$.

From the equations of the time evolution of the coil $I_L$ and induced
eddy $I_e$ currents, it is not difficult to notice that the transient
behaviour of those currents is characterized by the decay constants 
\begin{equation}
 \frac{1}{\tau_\pm}=-\frac{1}{2(1-k^2)}(\frac{1}{\tau}+\frac{1}{\tau_m})
 \pm\frac{1}{2(1-k^2)}
 \sqrt{(\frac{1}{\tau}+\frac{1}{\tau_m})^2-4(1-k^2)\frac{1}{\tau
     \tau_m}}\ ,
\label{decays}
\end{equation}
where $\tau=L/r$ is the Lenz effect time constant, $\tau_m=L_m/R_m$ is
the decay constant associated to the metallic surrounding, and $k$ is
a coupling parameter defined by the mutual inductance 
\mbox{$M^2=k^2 L L_m$}. Consequently, any measured magnetic field transient
can be mathematically described by a superposition of two exponential
terms with time constants defined by the quadrupole coil parameters and the
properties of the surrounding metallic environment. This means that one
way for checking the compensation of eddy--current--induced magnetic
field transients will be to reduce the double exponential behaviour of
the measured transient to a single one. As we will see later, the
designed driver can compensate not only the effect of eddy currents
but also speed-up magnetic field transients.

In terms of safety, there are two relevant factors to be considered in
order to design a current driver for switching the MOT coils. One is
the back induced $emf$ when the current is switched off, and the other 
is the maximum stored magnetic energy. In our case, for a resulting
inductance $L$ of the coils of 393 $\mu$H (wire resistance $r=$0.7
$\Omega$) the induced $emf$ is
\begin{equation}
 emf = -L\frac{dI_L}{dt} \approx -60\ V \ ,
\label{fem}
\end{equation}
for 15 A current change in about 100 $\mu$s. Concerning the stored magnetic energy $E$, for 
15 A of current we get
\begin{equation}
 E = \frac{1}{2}L I_L^2 \approx 44.2\ mJ \ .
\label{nrg}
\end{equation}

\begin{figure}[ht]
\begin{center}
\includegraphics*[width=1.0\textwidth]{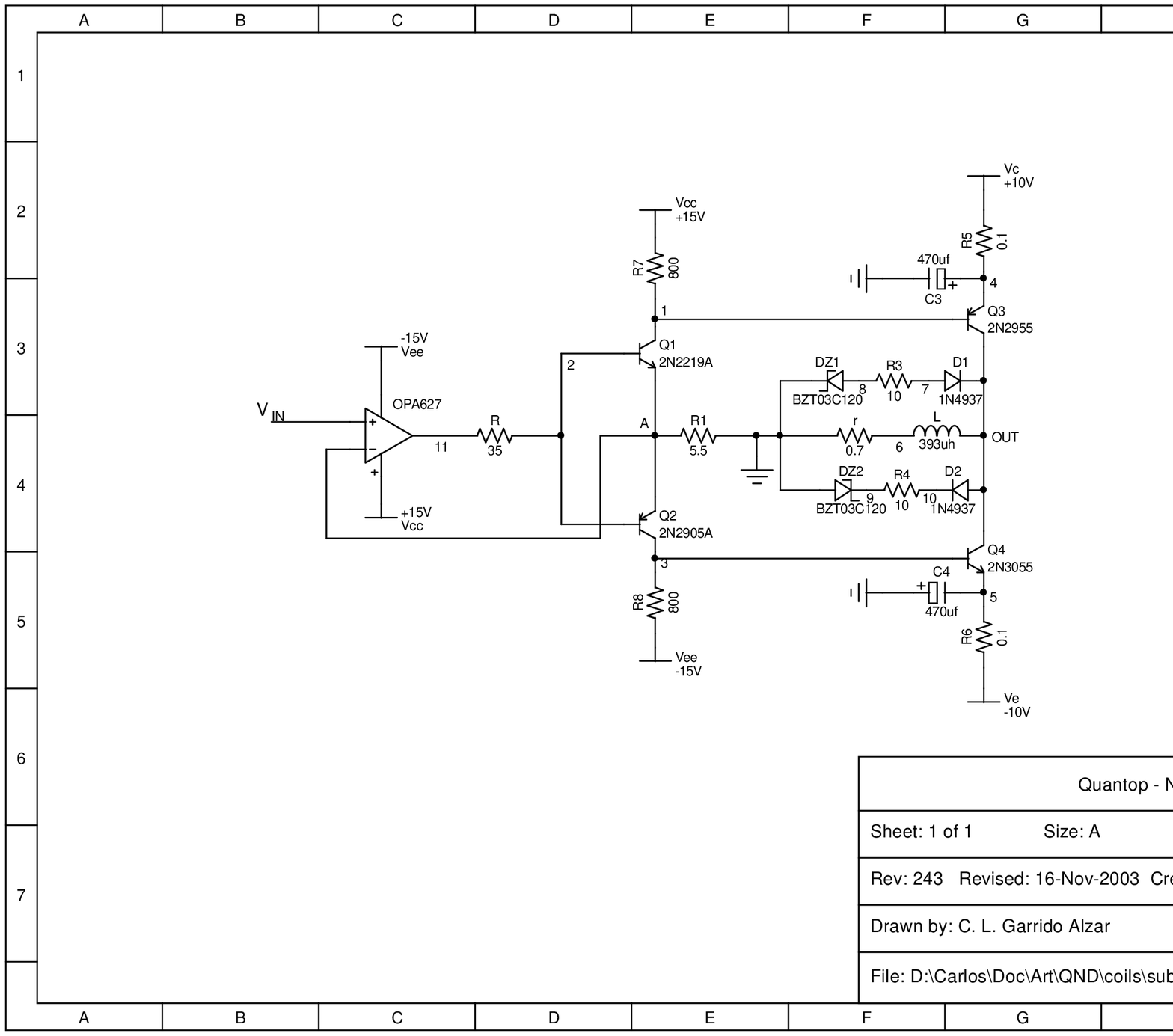}%
\end{center}
\caption{Bipolar current supply for fast switching of the magnetic field.}
\label{fig:driver}
\end{figure}

In the Fig.~\ref{fig:driver} the current driver designed to switch the
MOT coils is presented. The current supply for the coil is implemented using the power bipolar 
transistors 2N2955 and 2N3055, intended for power switching circuits. The operation of the 
power transistors is controlled by the control pulse $V_{IN}$ and the push-pull emitter 
follower stage realized by the high-speed switching transistors 2N2219 and 2N2905. These 
transistors, together with the operational amplifier, provide the necessary isolation of the 
control pulse source from the high power stage of the circuit.

During the transitions, the power transistors must be able to handle the developed $emf$ and 
therefore, we need to choose them with an emitter-collector breakdown voltage of at least the 
back induced voltage. In case of failure of the power transistors,
usually the collector is short to the base and, in that situation, it
is up to the switching transistors to handle the back 
induced $emf$. For this reason, we need to choose switching transistors with a collector-base 
breakdown voltage, again, of at least $emf$ to protect the control electronics. Apart from 
these considerations, the diodes $D1$ and $D2$ provide a decaying path for the energy 
accumulated in the coils and another piece of protection for the low voltage 
electronics~\footnote{For ultimate protection an optocoupler can be properly added}.

When the current through the coils is switched from some value to
zero, the coils alone induce a transient with a decay constant
$\tau=L/r$ of around 0.6 ms (Lenz effect). This time constant can be reduced 
by adding a resistor in the decaying path of the current. In the present circuit, 
we have added a 10 $\Omega$ resistor, reducing the decay constant in that way to 
36.7 $\mu$s. On the other hand, the use of the Zener diodes $DZ1$ and
$DZ2$ increases the protection of the circuit. They limit the 
voltage across the coils during the transitions and transform the long exponential decay due 
to $D1$ and $D2$ into a linear decay. Now, let's have a closer look into the functioning of 
this circuit.

\subsubsection*{Positive polarity of the control voltage}
When the control input voltage $V_{IN}$ is positive, $Q1$ is forward biased while $Q2$ is 
cut off. In this situation, the current through $R1$ is equal to the emitter current of $Q1$
\begin{equation}
 i_{E1} = -\frac{V_{IN}}{R1}\ ,
\label{eq1}
\end{equation}
and therefore, the current flowing into the base of $Q1$, with a dc current gain $\beta_1$, 
is 
\begin{equation}
 i_{B1} = \frac{V_{IN}}{R1(1+\beta_1)}\ .
\label{eq2}
\end{equation}

Since the base current necessary to put the power transistor $Q3$ into conduction equals the 
collector current of $Q1$ then, the current that flows through the collector of $Q3$, with a 
dc current gain $\beta_3$, is
\begin{equation}
 i_{C3} = -\frac{\beta_1 \beta_3}{R1(1+\beta_1)} V_{IN} \ .
\label{eq3}
\end{equation}

At this point, we know that $Q3$ is conducting. So, the diodes $D1$ and $D2$ are 
reverse and forward biased, respectively. Nevertheless, the voltage at the node OUT is 
inferior to the Zener voltage of $DZ2$ and consequently, very little current flows through 
$R4$. This means that the current in the coils $I_L$ is the same as the 
current in the collector of $Q3$, namely
\begin{equation}
 I_L = -\frac{\beta_1 \beta_3}{R1(1+\beta_1)} V_{IN} \ .
\label{eq4}
\end{equation}

\subsubsection*{Negative polarity of the control voltage}
In case of negative control voltage, the analysis proceeds as above with the appropriate 
dc current gains of the transistors in the final expression of the coils' currents.

\subsubsection*{Zero control voltage}
Let's suppose now that the voltage at the input node IN is zero. In this situation the 
current through $R1$ is zero since its terminals are at the same voltage level. Therefore, 
no current flows in the $Q1$ and $Q2$ collectors and consequently, the collector currents 
of $Q3$ and $Q4$ are also zero and so, no current flows into the coils.

Finally, from the expression of the coils' current, Eq.(\ref{eq4}), it can be seen that the 
level of this current can be adjusted, or controlled, by the applied voltage $V_{IN}$ at the 
input node of the circuit, an important feature for applications where adiabatic switching of 
the magnetic field gradient is required. Moreover, having fixed the resistor $R1$ and the dc 
current gains of the switching transistors, it is possible to increase the current limit by 
choosing power transistors with different $\beta_3$ ($\beta_4$).

From the preceeding analysis it is clear that the other sensitive elements of this circuit are 
the resistors $R$ and $R1$, chosen as follows. Assuming a reduction of $\beta_3$ 
($\beta_4$) from 20 to 5 for a peak coils' current of 10 A, we will need at the base 
of $Q3$ a current of 2 A to maintain it into conduction. Therefore, for a maximum input 
control voltage of 10 V the resistor $R1$ is 5 $\Omega$. On the other hand, to find the 
specification of the base resistor $R$, we demand the output voltage of the operational 
amplifier to be 2 or 3 \% below its supply voltage, $\pm$15 V in our case. This yields the 
following criterion for $R$
\begin{equation}
 R \le \frac{12 - V_{IN} - V_{BE1}}{I_{B1}}\ .
\label{rcrit}
\end{equation}

If we take the dc current gain $\beta_1$ of the switching transistor $Q1$ to be 70, 
we obtain a peak base current $I_{B1}$ of 28.6 mA and, for 10 V of control voltage, 
$V_{BE1} = 1$ V, we get for our designed circuit that $R \le 35\ \Omega$.

\section{Simulations and experimental results}
\begin{figure}[htb]
\begin{center}
\includegraphics*[width=0.7\textwidth]{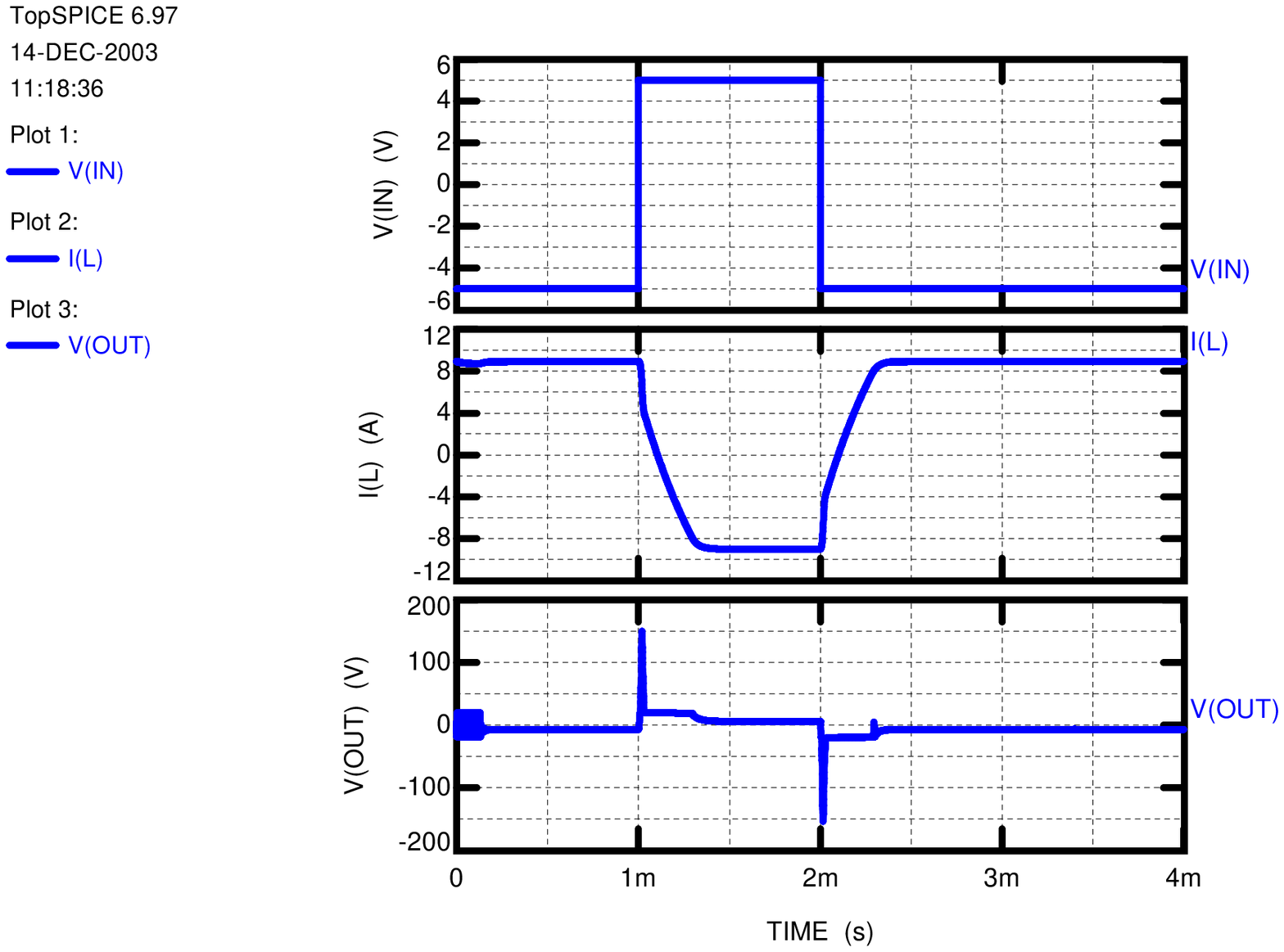}%
\end{center}
\caption{Current and back emf responses of the current driver.}
\label{fig:out1}
\end{figure}

The behaviour of the current driver circuit in Fig.~\ref{fig:driver} has been simulated using 
\mbox{TopSPICE 6.97~\cite{penzar}} and the results of those simulations are presented in 
Fig.~\ref{fig:out1}. In this figure we show the coils' current and 
back induced $emf$ transient responses of the circuit to a control voltage pulse of 1 ms 
duration. The input voltage level was changed by 10 V (from -5 V to +5 V).

As we can see in Fig.~\ref{fig:out1}, the current flowing through the coils, at the beginning
$\sim$9 A, decays to zero very fast (in around 100 $\mu$s). The initial fast almost linear 
decay is dominated by the effect of the resistor $R4$. Then, the current decay is 
controlled by the action of the Zener diode $DZ2$. The total transition time for reversing the 
current polarity from +9 A to -9 A is thus around \mbox{340 $\mu$s}. For these control voltage 
and current settings, the back induced voltage during the transitions is close to 150 V.

Since the switching time of the magnetic field depends on how much current was flowing in the 
coils before the transition started, we also performed simulations for lower input control 
voltages. When the input voltage was changed from -0.5 V to +0.5 V, total switching times of the 
order of \mbox{100 $\mu$s} are obtained, with the corresponding reduction in the back induced 
$emf$.

\begin{figure}[htb]
\begin{center}
\includegraphics*[width=0.8\textwidth]{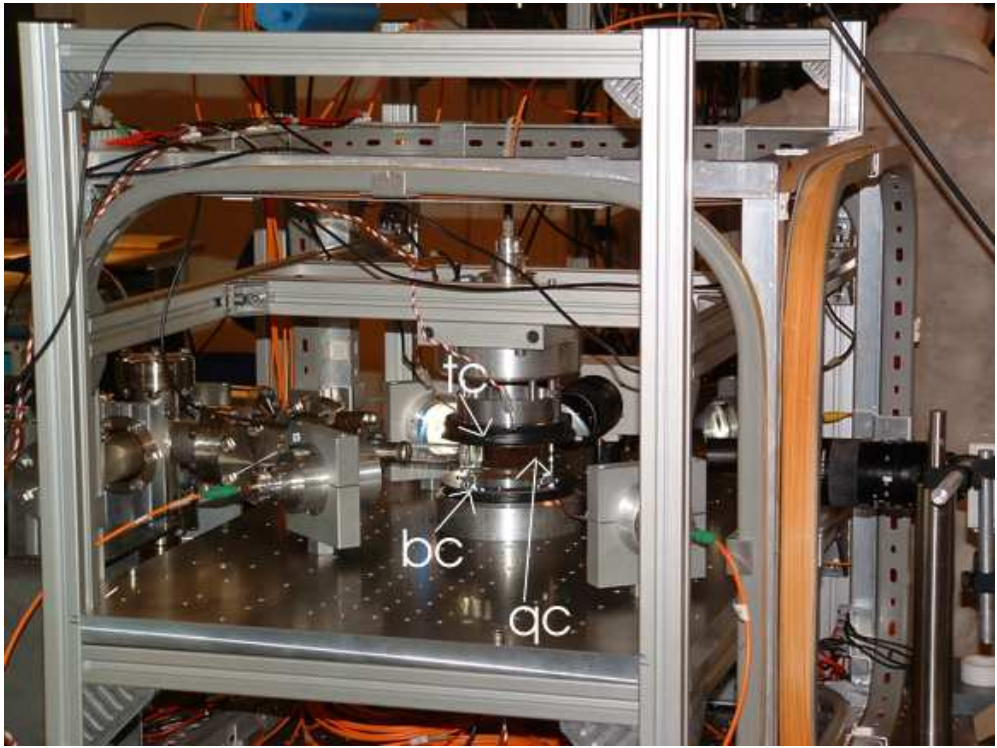}%
\end{center}
\caption{Experimental setup. When measuring the magnetic flux
  transients associated with eddy currents, the pick-up coil is placed in
the position occupied by the quartz cell (qc), between the top (tc)
and bottom (bc) gradient coils (black rings). At this point the
magnetic fields produced by the tc and bc coils cancel and we are more
sensitive to any magnetic flux generated by eddy currents.}
\label{fig:set}
\end{figure}

So far the analyzed results concerned the behaviour of the current flowing in the 
coils. However, the switching characteristics of the magnetic field itself are different because 
of the effect of eddy currents. For that reason, our next step was the study of magnetic 
field transients when the coils are driven by the circuit designed in the last section. We 
measured the magnetic flux transient using a small (compared to the
MOT coils) critically damped pick-up coil ($Lp=$ 7 $\mu$H inductance 
and 0.1 $\Omega$ wire resistance) connected in parallel with a capacitor 
(\mbox{$C=$ 2.2 $\mu$F}) and in series with a 3.3 $\Omega$
resistor. The capacitor limits the frequency bandwidth 
of the sensing circuit making the measured magnetic flux rate less susceptible to high frequency magnetic field 
noise generated by, e.g., radio stations, and the resistor value is
chosen in order to obtain critically damped oscillations when the pick-up coil is
kicked by a fast magnetic flux change. We placed the pick-up coil midway between the MOT 
gradient coils where the atomic cloud is normally formed, that is to
say in the position occupied by the quartz cell (qc) in the Fig.~\ref{fig:set}. This location was chosen because 
it is at this point that the magnetic field produced by the MOT coils is zero and consequently, 
by symmetry considerations, any detectable variation of the magnetic field there is directly 
related to eddy currents. It can be clearly seen from the
picture that the two gradient coils (tc and bc) are surrounded by different
metallic components, being the bottom coil on an aluminium plate. When
the magnetic field produced by these coils is switched off, it will
create eddy currents with a spatial distribution, generating an
unwanted magnetic field gradient.

\begin{figure}[htb]
\begin{center}
\includegraphics[width=0.8\textwidth]{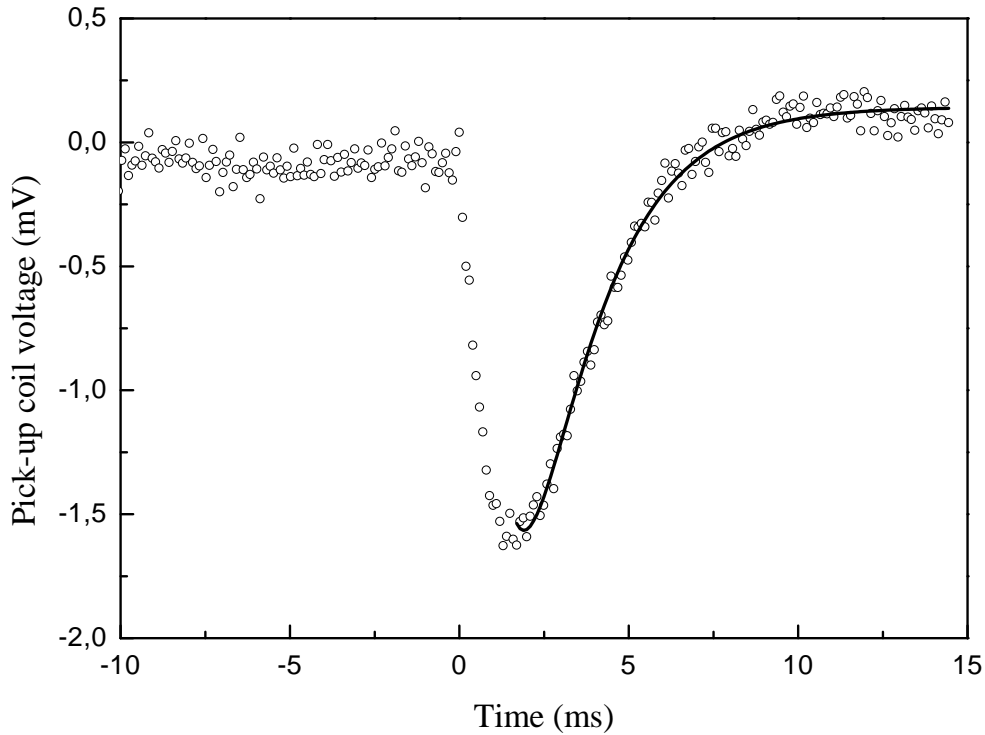}%
\end{center}
\caption{Pick-up coil voltage for an initial nominal current in the
  coils $I_{L}=3.1$ A. The two exponential fit (solid line) gives
  decay constants equal to 1.97(17) ms and 0.80(18) ms.}
\label{fig:mea1}
\end{figure}

As a reference measurement, we used a commercial current supply with switching times on the 
order of hundreds of microseconds to drive the current $I_L$ in the
quadrupole coil. The transient response when the current was 
switched off from an initial value of 3.1 A is shown in Fig.~\ref{fig:mea1}. As it can 
be seen in that figure, the duration of the measured settling time was around 10 ms, much 
larger than the power supply switching time and therefore, attributed to the influence of 
magnetic field transients due to eddy currents. Together with the
experimental points, is also shown in Fig.~\ref{fig:mea1}
the best theoretical fit (solid line), given by a superposition of two
exponentials with time constants of 1.97(17) ms and
0.80(18) ms, respectively. Since these characteristic times are greater
than the expected value corresponding to the Lenz effect alone 
(36.7 $\mu$s), we conclude that we have an important influence
of eddy currents in the measured magnetic field transient.

\begin{figure}[htb]
\begin{center}
\includegraphics[width=0.8\textwidth]{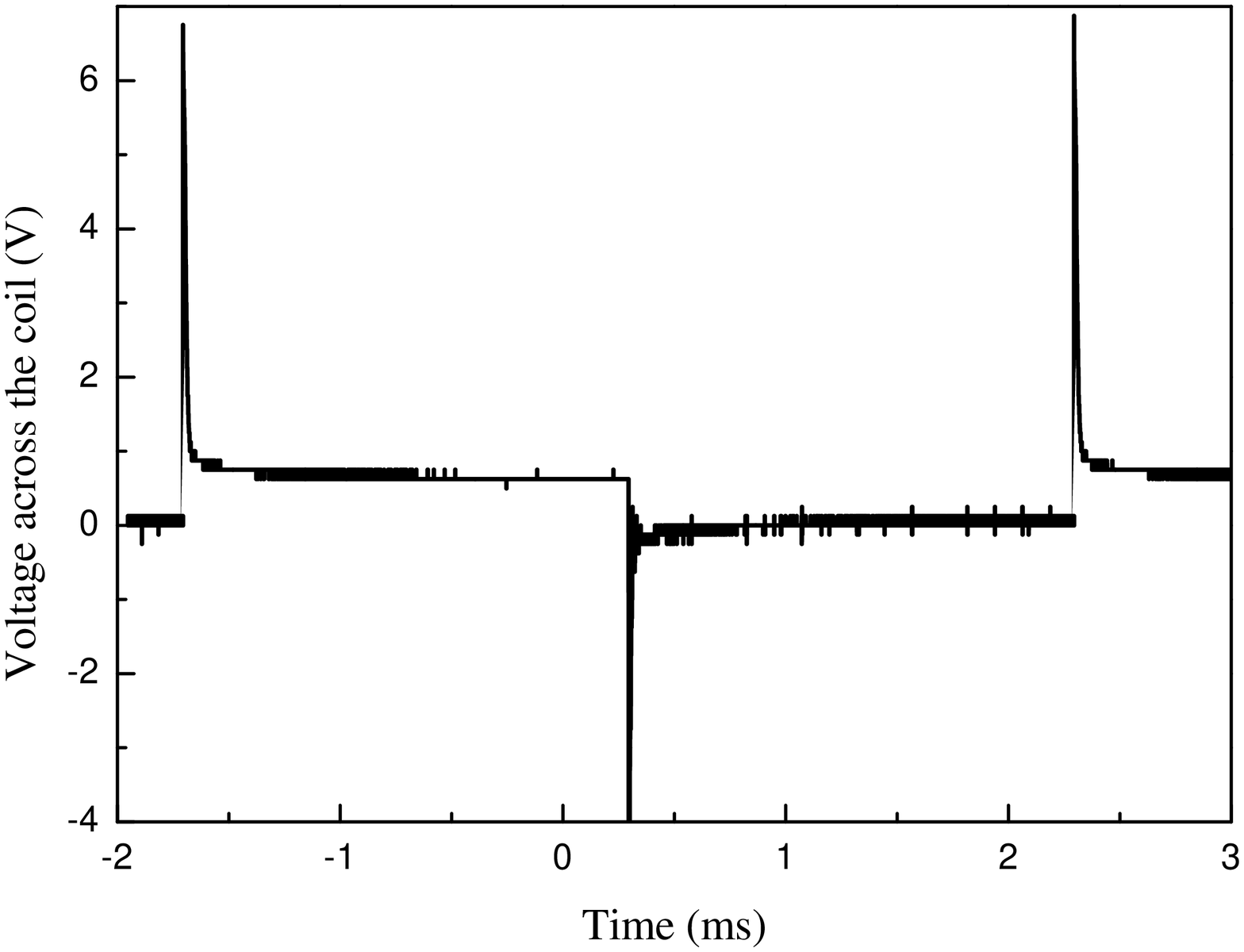}%
\end{center}
\caption{Transient response of the voltage across the MOT coils.}
\label{fig:mea2}
\end{figure}

The first test of the designed circuit (Fig.~\ref{fig:mea2}) was the
measurement of the switching waveform corresponding to the voltage across the MOT coils, a quantity equivalent to
the simulated $V(OUT)$ in Fig.~\ref{fig:out1}. The first voltage spike in
this trace was obtained when the coil current was switched on,
to 6.6 $A$. For the second one, the current was
reversed from 6.6 $A$ to -3.7 $A$, before setting it to 0. 

\begin{figure}[htb]
\begin{center}
\includegraphics[width=0.7\textwidth]{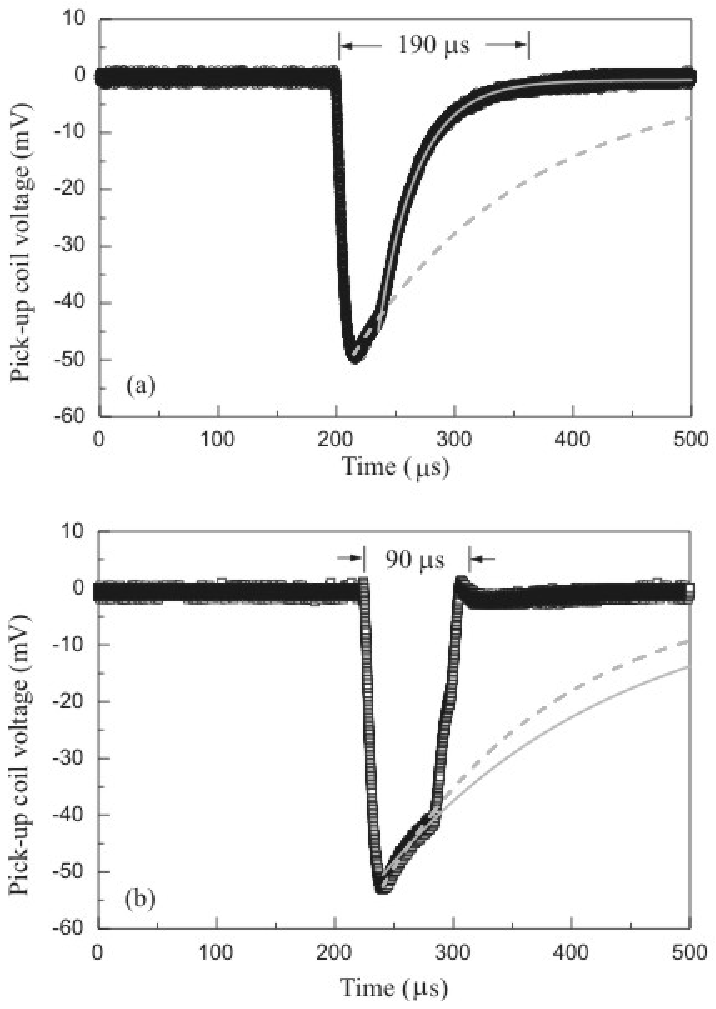}%
\end{center}
\caption{(a) Compensation of the magnetic field transient induced by eddy
  currents (dashed line). The decay constant, obtained from the single
  exponential fit (solid line) to the Lenz effect transient, was
  32.57(2) $\mu$s. (b) Transient speed-up by compensation of the Lenz
  effect.}
\label{fig:mea3}
\end{figure}

As expected, the duration of the transients were on the microsecond time
scale. However, in order to draw any conclusion about the compensation
of the effects due to eddy currents, a detailed look at the magnetic
flux transient is required. This is done in Fig.~\ref{fig:mea3}(a) and
(b). These measurement were taken when the current in the MOT coil,
initially 6.6 $A$, was briefly reversed to -3.7 $A$ before setting it
to zero (as in the middle voltage spike in Fig.~\ref{fig:mea2}).

Initially, we tested the compensation of the magnetic flux induced by
eddy currents. The result of this measurement is presented in
Fig.~\ref{fig:mea3}(a), obtained when the Zener diodes $DZ1$ and $DZ2$
are each replaced by a short and therefore, we allow more current to
flow through the protection diodes $D1$ and $D2$ when the back
induced $emf$ appears. In this figure, we attribute the initial exponential decay
(dashed line) with a time constant of 150 $\mu$s to eddy
currents. This decay is shortened down to around 40 $\mu$s when the
current is reversed, leaving the single exponential behaviour (solid
line) characterized by a time constant of 32.57(2) $\mu$s, in good
agreement with the modified time constant of the Lenz effect. The
overall settling time of the transient is about 190 $\mu$s.

Since with the designed driver we are able to provide a larger $dI_L/dt$ compared to the one
given by the commercial power supply, we induce larger eddy currents
as it can be seen from the absolute maximum reached by the pick-up
coil voltage. Nevertheless, the eddy--current--induced magnetic field
transient is compensated by the reversal of the coils current.

In the next test, we try to reduce the settling time by
reducing the Lenz effect decay. This compensation is achieved by
removing the short across the Zener diodes. In this case, a current
will flow through the protection diodes \textit{only} when the voltage at the
node OUT in Fig.~\ref{fig:driver} will be larger then the respective
Zener voltage. Consequently, a larger amount of the compensating current from the
power supply will flow through the quadrupole coil. This effect
is shown in Fig.~\ref{fig:mea3}(b), where the time constants for the
induced eddy current magnetic flux transients, solid and dashed lines,
are 200 $\mu$s and 150 $\mu$s, respectively. The settling time, or the
overall transient duration, is now around 90 $\mu$s.

As it was mentioned in the introduction, the magnetic field transients
induced by eddy currents can wash out the samples of cold atoms
prepared in the MOT. Therefore, we used this fact in order to perform
our ultimate test of the driver. Using a sample of cold caesium atoms
(at around $120\ \mu$K), we realized an experiment similar to a
previous one presented in~\cite{moi}. In short, we built a
Mach--Zehnder interferometer around the cell containing the cold atoms
and we measured the phase shift that a probe laser pulse, with around
$10^{7}$ photons, experience after interaction with the atoms. The
laser was red detuned by 25 MHz from the Cs cycling transition 
$6S_{1/2}(F\!=4) \rightarrow 6P_{3/2}(F'\!=5)$. The used protocol for
this measurement was as follows: 
\begin{itemize}
 \item we trap the atoms during 3 seconds;
 \item the magnetic quadrupole field, the trapping and the repumper lights are
       turned off during 10 ms and, we fire a pulse train with 100
       probe pulses (pulse duration 2 $\mu$s, repetition period 6 $\mu$s) that
       interact with the atoms;
 \item during 8 ms we turn on the trapping light only to remove the atoms from the
       probing region;
 \item we take a reference measurement with another sequence of 100 pulses.
\end{itemize}

\begin{figure}[htb]
\begin{center}
\includegraphics[width=0.8\textwidth]{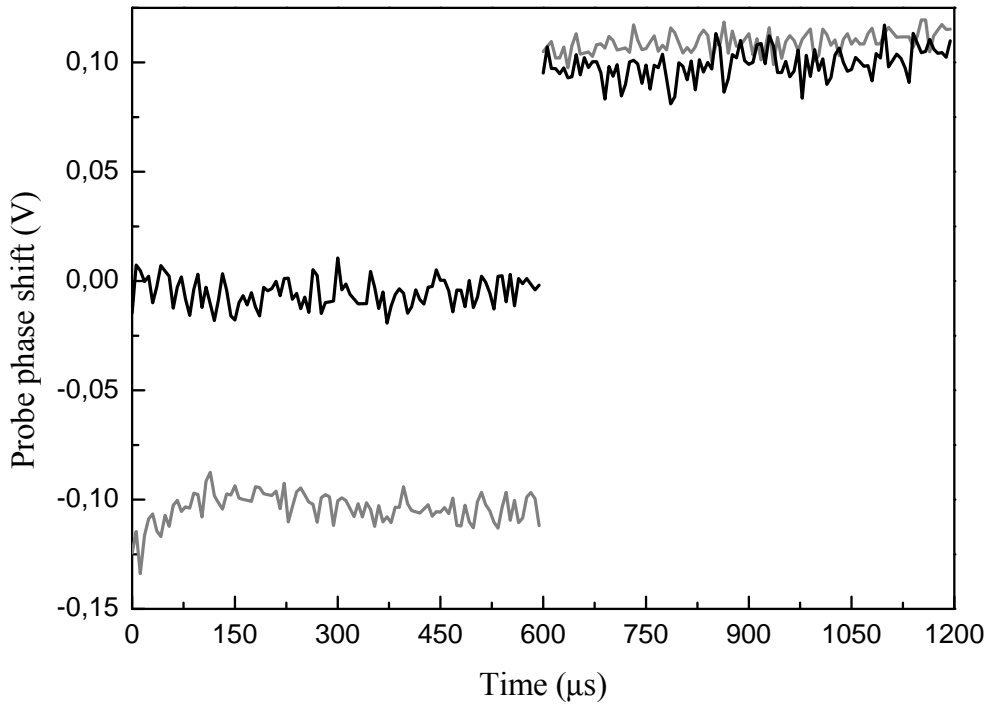}%
\end{center}
\caption{Phase shift experienced by laser pulses after the
  interaction with a sample of cold atoms. The gray (black) traces
  correspond to the situation in which the current in the MOT coils
  is switched using the driver (commercial unit).}
\label{fig:phase}
\end{figure}

The above steps are repeated 50 times in order to obtain a reliable
data and the results are presented in Fig.~\ref{fig:phase}. In one
case the experiment is done switching the quadrupole magnetic field
with the commercial unit and in the other with the driver.

The traces covering the first 600 $\mu$s represent the probe phase shift
after the interaction with the atoms. The results of the reference
measurement, probe phase shift in the absence of atoms, cover the
interval starting from 600 $\mu$s up to 1200 $\mu$s. As it can be seen
from the Fig.~\ref{fig:phase}, the measured atomic contribution when the
quadrupole magnetic field of the MOT is controlled by the driver (gray
traces) is larger than the atomic contribution measured when the control is done by the
commercial unit (black traces) and without the possibility of compensation of
magnetic field transients due to eddy currents. Since the reference
measurement results are at the same level in both cases, we can without
ambiguity attribute the reduction in the signal to the fact that less
atoms are probed by the laser pulses.

Finally, from Fig.~\ref{fig:mea3} it is clear that the magnetic field
can be switched off about 100 times faster than in the case without
polarity change of the coils' current. This dramatic
reduction of the transition time from the millisecond down to the
microsecond time scale, demonstrate once again that one of the key
factors in the compensation of the undesirable eddy--current--induced
magnetic field transients is the possibility of changing very fast the
polarity of the current in the MOT quadrupole coils. In that way, we can overshoot 
or undershoot the current during the transitions, providing the necessary magnetic flux for the 
compensation of stray magnetic fields.

\begin{acknowledgments}
This research has been supported by the Danish National Research Foundation and by the CAUAC 
European research network.
\end{acknowledgments}


\end{document}